\documentclass[smallextended]{svjour3}       
\smartqed  

\usepackage[numbers]{natbib}
\usepackage{graphicx}

\begin{document}

\title{Wave Fronts in a Causality Violating G\"{o}del-Type Metric}

\author{Thomas P. Kling       \and
       Faizuddin Ahmed \and 
        Megan Lalumiere
}


\institute{T. Kling \at
              Dept. of Physics, Bridgewater State University,
Bridgewater, MA 02325 \\
              Tel.: +1-508-531-2895\\
              Fax: +1-508-531-1785\\
              \email{tkling@bridgew.edu}           
           \and            
              F. Ahmed \at
         Ajmal College of Arts and Sciences, Dhubri-783324, Assam, India\\
         Tel: +91-863-830-2021\\
         \email{faizuddinahmed15@gmail.com}  
         \and
            M. Lalumiere \at
              Dept. of Physics, Bridgewater State University,
Bridgewater, MA 02325 \\
              \email{mlalumiere@student.bridgew.edu} 
}

\date{Received: date / Accepted: date}

\maketitle

\begin{abstract}

\noindent The light-rays and wave fronts in a flat class of G\"{o}del-type metric are examined to reveal the causality violating features of the space-time.  Non-causal features demonstrated by the development of unusual wave front singularities are shown to be related to the non-monotonic advance of time along the light rays, as measured by a system of observers at rest with respect to one another with synchronized clocks.

\end{abstract}


\maketitle


\section{Introduction}

In general relativity, locally space-times have the same causal structure as flat, Minkowski space-times, because all space-times are locally Minkowskian. On the non-local scale, however, significant differences may arise as the field equations of general relativity do not provide non-local constraints on the underlying space-times. Indeed, it has long been known that there are solutions to the Einstein's field equations that present non-local causal anomalies in the form of closed time-like curves, closed null geodesics and closed time-like geodesics.

The renowned model found by G\"{o}del \cite{Godel} is possibly the best known example of a solution to the Einstein's field equations, with a physically well-motivated source, possessing causal anomalies. The G\"{o}del solution makes it apparent that general relativity permits solutions with closed time-like worldlines, even when the metric possesses a local Lorentzian character that ensures an inherited regular chronology, and therefore, the local validation of the causality principle. 

This modern understanding of the G\"{o}del metric has led to an increased interest in examining solutions to the Einstein's field equations with unusual topological features, all of which suggest a broad richness in possible theories of gravity. Solutions have examined that have closed time-like curves (see {\it e. g.}, \cite{EPJC} and related references there in), closed null geodesics ({\it e. g.}, \cite{AOP}), and closed time-like geodesics ({\it e. g.}, \cite{PTEP} and related references their in) in space-times with unusual equations of state. Section~\ref{character:sec} describes a range of class of the G\"{o}del style metrics, including the one analyzed in more detail in this paper, and enumerates the current understanding of geodesics in those metrics.

To gain broader insight into these anomalies, we examine the properties of light rays and wave fronts of null geodesics emitted by point sources in a G\"{o}del-type metric in this paper. The light rays and null wave fronts reveal some of the causality violating features of the space-time, and we contrast the structure of these wave fronts with those in the G\"{o}del space-time. These comparisons allow us to both discuss the nature of wave fronts as non-local phenomenon and discuss the causality violating aspects that are possible. Because not much is known about the broad properties of these new G\"odel-type metrics, this paper provides an unique window into some of their properties.

Section \ref{character:sec} discusses the variety of G\"{o}del-type metrics possible in order to position the metric considered in this paper.  In Section ~\ref{equations:sec}, we outline the equations for geodesics and discuss the meaning of the time variable which will be key in our understanding of wave fronts.  Sections~\ref{lightrays:sec} and ~\ref{planefronts:sec} present pictures of individual light rays and the wave fronts and discuss how these develop in time for rays in plane of symmetry.  A more thorough discussion of how wave front singularities, and other non-causal features, arise is discussed in Section~\ref{lightcones:sec}.  In Section~\ref{threeDfronts:sec}, three dimensional wave fronts are presented, followed by a final discussion and conclusion.

\section{Characterization of a Linear Class of G\"{o}del-type Metrics }\label{character:sec}

A particularly simple, Type D metric has recently examined that is a member of a linear class of G\"{o}del-type metrics \cite{ahmed1}. Previously, relativistic quantum motion of spin-$0$ particle without interactions \cite{ahmed2}, linear confinement of a scalar particle subject to a scalar and vector potentials of Coulomb-types \cite{ahmed3}, Dirac equation \cite{ahmed4}, spin-$0$ system of DKP equation \cite{ahmed5}, and DKP oscillator \cite{ahmed6} was investigated in this Type D metric.

In cylindrical coordinates $(t,r,\phi,z)$, the general class of G\"{o}del-type metrics are given by \cite{ahmed4,Hassan,Reboucas,JS}:
 \begin{equation}
     ds^2 = - [dt + H(r) d\phi ]^2 + dr^2 + D^{2}(r)d\phi^2  +dz^2.
     \label{1}
 \end{equation}
 The necessary and sufficient conditions for the G\"{o}del-type metric (\ref{1}) to be space-time homogeneous (ST-homogeneous) are given \cite{ahmed4,Reboucas,JS,Reboucas2}:
 \begin{equation}
     \frac{H'}{D} = 2\,\Omega, \quad \frac{D''}{D} = \mu^2,
     \label{2}
 \end{equation}
\noindent where $\Omega^2 > 0$ and $-\infty \leq \mu^2 \leq +\infty$ are constants, prime denotes derivative with respect to $r$.

There are several different class of G\"{o}del-type space-time geometries, including a linear;  trigonometric or spherical, and hyperbolic class which we discuss below.

\vspace{0.1cm}
{\bf Case 1}: Linear class:
\vspace{0.1cm} 

To obtain a linear class of G\"{o}del-type metrics, one can takes $\mu =0$ in the condition  Eqn.~\ref{2}. The Som-Raychaudhuri \cite{Som} metric and the solution given by Reboucas {\it et al.}  \cite{Reboucas} are the examples of this class of solutions. In our case, the study space-time satisfies the condition Eq. (\ref{2}) with $\mu=0$ which clearly indicates that the metric we consider in this paper also belong to this linear class metrics.

The Som-Raychaudhri metric in polar coordinates $(t,r,\phi,z)$ is given by
\begin{equation}
    ds^2=-(dt+\Omega\,r^2\,d\phi)^2+r^2\,d\phi^2+dr^2+dz^2
    \label{3}
\end{equation}

\vspace{0.1cm}

{\bf Case 2}: Spherical class:
\vspace{0.1cm}

With $\mu^2 < 0$ in Eqn.~\ref{2}, in a G\"{o}del-type space-time, the space-time belongs to trigonometric, or spherical, class of G\"{o}del type metrics (see \cite{Reboucas}).

\vspace{0.1cm}

{\bf Case 3}: Hyperbolic class:
\vspace{0.1cm}

On the other hand, with $\mu^2>0$ in Eqn.~\ref{2}, the space-time belongs to the hyperbolic class of G\"{o}del-type metrics (see {\it e. g.} \cite{Reboucas}).

The geodesic equations of motion in G\"{o}del-type space-times have analyzed by several authors. Numerous publications concerning the solutions to the geodesic equations of G\"{o}del universe and Som-Raychaudhuri space-time, were known in the literature dating back at least to \cite{Paiva}. The geodesic equations for the G\"{o}del metric was firstly solved by Kundt \cite{Kundt} where, the Killing vectors and corresponding constants of motion were used. Novello {\it et al.} \cite{Novello} have studied a detailed discussion on the geodesic motion in the G\"{o}del universe using the method of effective potential as well as the analytical solution. Pfarr \cite{Pfarr} investigates both geodesic and non-geodesic motion of test particles in the G\"{o}del universe. More recently, a very convenient set of coordinates, in which the G\"{o}del universe reveals its properties more apparently, were discussed \cite{Kajari}. There the symmetries of G\"{o}del universe were provided, alongside a special solution to the geodesic equations. The understanding of null geodesics provides the bedrock for ray tracing in G\"{o}del and G\"{o}del-type space-time geometries. Egocentric visualizations of certain scenarios in G\"{o}del universe can be found in Grave {\it et al.} \cite{Buser}. There, the authors presented much more improvements for visualization techniques regarding general relativity. Furthermore, finite isometric transformations were used to visualize illuminated objects. This method was technically reworked and improved in \cite{Grave} and resulted in an interactive method for visualizing various aspects of G\"{o}del universe from an egocentric perspective. There, an analytical solution to the geodesic equations and a numerical integration of the equations of isometric transport, were used.  The analytical solution of the geodesic equations of G\"{o}del's universe for both particles and light in a special set of coordinates which reveals the physical properties of this space-time in a very transparent way was studied in \cite{FG}. They also recapitulate the equations of isometric transport for points and derive the solution for G\"{o}del's universe. The G\"{o}del universe through worldlines associated with motion at constant speed and constant acceleration orthogonal to the instantaneous velocity (WSAs) was studied in \cite{DS}. They have shown that these world-lines can be used to access every region - both spatial and temporal - of the space-time.

The approaches of previous authors, in particular F.M. Paiva {\it et al.} \cite{Paiva} to solving for individual geodesics are similar to the results of this paper. The current study differs by considering families of null geodesics, emitted at one space-time point in all spatial directions, and organizing the
points along the geodesics according to a time foliation consistent with the observations of the observer who emitted the light pulse. The global structure of the null geodesic wave fronts organized in this fashion shed a different perspective on the time anomalies described in the literature when looking at individual geodesics.

\section{Geodesics for Observers and Light Rays} \label{equations:sec}

In this paper, we analyze a particular G\"odel-type space-time of Type D metric, a  linear or flat class of G\"{o}del-type solutions \cite{ahmed1}. This metric is given by (see also, Refs. \cite{ahmed2,ahmed3,ahmed4,ahmed5})

\begin{equation}
    ds^2 = - dt^2 + dx^2 + (1-\alpha_o^2 x^2) dy^2 - 2 \alpha_o x dt dy + dz^2 \label{metric}
\end{equation}

\noindent where $\alpha_o>0$ is a real parameter and the Cartesian coordinates extend from minus to plus infinity. 

Applying the condition in Eqn.~\ref{2} into the metric Eqn.~\ref{metric} we get
\begin{equation}
    \alpha_o=2\,\Omega,\quad \mu=0.   \label{metric-condition}
\end{equation}
which clearly indicates that our considered space-time  belongs to the class of flat G\"{o}del-type metrics.

Based on the metric as presented in Eqn.~\ref{metric}, we introduce a Lagrangian defined by 

\begin{eqnarray} \mathcal{L} &=& \frac{1}{2} g_{ab} \dot x^a \dot x^b \nonumber \\ ~& = & ~ -\dot t^2 + \dot x^2 +  (1-\alpha_o^2 x^2) \dot y^2 - 2 \alpha_o x \dot t \dot y + \dot z^2 \label{lagrangian}\end{eqnarray}

\noindent where $\cdot$ represents a derivative with respect to an affine parameter $s$. Because $t$, $z$ and $y$ do not appear in the Lagrangian, we have three conserved momentum $p_t$, $p_y$ and $p_z$.  This means that the Euler-Lagrange equations will result in one second-order, and three first-order, ordinary differential equations for geodesics.

\subsection{Geodesic Equations}
The Euler-Lagrange equations for $y$ and $t$ are trivially coupled:

\begin{eqnarray}  - p_t & = &  -\dot t - \alpha_o x \dot y \nonumber \\ p_y & = & (1-\alpha_o^2 x^2 ) \dot y - \alpha_o x \dot t \label{originalyt}. \end{eqnarray}

\noindent In the $t$ Euler-Lagrange equation, we choose the constant as $-p_t$ so that at an initial condition where $\dot y = 0$, positive $p_t$ values correspond with positive $\dot t$. Equations~\ref{originalyt} can be trivially decoupled, resulting in 

\begin{eqnarray}
\dot t & = & (1-\alpha_o^2 x^2 ) p_t - \alpha_o x p_y \label{tdot} \\ \dot y & = & p_y + \alpha_o x p_t. \label{ydot} 
\end{eqnarray}

For the $x$ coordinate, the Euler-Lagrange equation results in

\[ \ddot x = -\alpha_o^2 x \dot y^2 - \alpha_o \dot t \dot y, \]

\noindent but we can substitute our results for $\dot y$ and $\dot t$, and simplify the result to obtain

\begin{equation} \ddot x  = - \alpha_o p_t p_y + \alpha_o p_t^2 x. \label{xddot} \end{equation}

Equation~\ref{xddot} is the equation for a simple harmonic oscillator with an offset to the central position from the origin.  Defining

\begin{equation} u = x + \frac{p_y}{\alpha_o p_t} \end{equation}

\noindent and substituting into Eqn.~\ref{xddot} results in 

\[ \ddot u = -\alpha_o^2 p_t^2 u, \]

\noindent so that the solution for $x(s)$ is given by 

\begin{equation} x = -\frac{p_y}{\alpha_o p_t} + A \sin(\alpha_o p_t s) + B \cos(\alpha_o p_t s) \label{xsolution} .\end{equation}.

\noindent With this solution for $x(s)$, we can solve for $y(s)$ by simply substituting into Eqn.~\ref{ydot}, with the result

\begin{equation} y = C - A \cos(\alpha_o p_t s) + B \sin(\alpha_o p_t s) \label{ysolution} . \end{equation}

\noindent The equation for $z(s)$ is trivially 

\begin{equation} z = p_z s. \end{equation}

\noindent A similar substitution procedure results in an analytic expression for $t(s)$, even though this expression is not simple.  

For convenience and future reference, we note that the solutions for the spatial part of the geodesics and the velocities associated with these coordinates is given by 

\begin{eqnarray}
x & = & -\frac{p_y}{\alpha_o p_t} + A \sin(\alpha_o p_t s) + B \cos(\alpha_o p_t s) \nonumber \\
\dot x & = & A \alpha_o p_t \cos(\alpha_o t s) - B \alpha_o p_t \sin(\alpha_o p_t s) \nonumber \\
y & = & C - A \cos(\alpha_o p_t s) + B \sin(\alpha_o p_t s) \nonumber \\
\dot y & = & A \alpha_o p_t \sin(\alpha_o p_t s) + B \alpha_o p_t \cos(\alpha_o p_t s) \nonumber \\ z & = & p_z s \nonumber \\ \dot z & = & p_z. \label{fullsolutions}  
\end{eqnarray}

\subsection{Boundary conditions for static observers}\label{static_obs:sec}

In this paper we wish to examine the wave fronts of null geodesics. Wave fronts are fundamentally non-local constructions, particularly wave fronts of constant time.  An observer must {\emph{reconstruct}} the wave front based on information from observers distributed throughout the space, who all, ideally, have synchronized clocks. 

From the metric, we notice that an observer at rest in the space-time would carry a clock which reads the proper time.  The question is whether an observer at the origin can work with a team of observers who remain at rest relative to her.  If so, then a system of synchronized clocks reading the proper time can be constructed, and a wave front of constant time is a physically realizable non-local construct.

Our goal therefore is to show that there are time-like geodesics for which the observers hold their position at $(x_o, y_o,z_o)$.  This is trivial in the $\hat z$ direction: $p_z = 0$.  From Eqn.~\ref{ydot} we see that for $\dot y = 0$ at all $s$, we need to fix the value of $p_y$:

\begin{equation} p_y = -\alpha_o p_t x_o. \label{py:static} \end{equation}

\noindent With this restriction on $p_y$, we see that $\ddot x$ is zero at $s = 0$,

\[ \ddot x = - \alpha_o p_t \, (-\alpha_o p_t x_o) - \alpha_o^2 p_t^2 x_o = 0.\]

\noindent  Further, Eqn.~\ref{py:static} implies that the constant in $x(s)$ equation is $x_o$.  Therefore, with the restriction on $p_y$ given in Eqn.~\ref{py:static} with $p_z=0$, we see that it is simply possible to set $A = B = 0$ in Eqns.~\ref{fullsolutions} with $C = y_0$.  The result is that the geodesic maintains the position $(x_o, y_o, z_o)$ with no velocity or acceleration.

\subsection{Boundary conditions for null geodesic wave fronts}

To consider wave fronts of null geodesics, we would now like to consider the boundary conditions to be of the form that at $s = 0$, the light rays are at the origin (with $t=0$) with initial velocities given by $(\dot t,\,\dot x,\, \dot y, \, \dot z) = (p_t, \, p_x, \, p_y, \, p_z)$.  We see that this can be accomplished by setting 

\[ B = \frac{p_y}{\alpha_o p_t}, \]
\[ A = C = \frac{p_x}{\alpha_o p_t}. \]

\noindent In this case, we now have

\begin{eqnarray} 
x & = & -\frac{p_y}{\alpha_o p_t} + \frac{p_x}{\alpha_o p_t} \sin(\alpha_o p_t s) + \frac{p_y}{\alpha_o p_t} \cos(\alpha_o p_t s) \nonumber \\
y & = & \frac{p_x}{\alpha_o p_t} - \frac{p_x}{\alpha_o p_t} \cos(\alpha_o p_t s) + \frac{p_y}{\alpha_o p_t} \sin(\alpha_o p_t s) \nonumber \\
z & = & p_z s. \label{nullgeodesics} \end{eqnarray}

To ensure that the geodesics are null geodesics, we need the Lagrangian, Eqn.~\ref{lagrangian}, to be zero at $s = 0$.  Using the expressions above in Eqn.~\ref{nullgeodesics}, this results in

\begin{equation} - p_t^2 + p_x^2 + p_y^2 + p_z^2 = 0. \end{equation}

\noindent We are free at this point to set $p_t = 1$.  Then, re-arranging the restriction that the geodesics are light rays results in 

\[ p_x^2 + p_y^2 = 1 - p_z^2, \]

\noindent so that we see that $p_z$ is restricted in the range $-1 \le p_z \le +1$. Introducing a parameter $\gamma$, we take

\begin{eqnarray}
p_x &=& \pm \sqrt{1 - p_z^2}\cos{\gamma} \nonumber \\ p_y &=& \pm \sqrt{1 - p_z^2} \sin \gamma \label{gamma}.
\end{eqnarray}

\noindent The meaning of $\gamma$ is that for a light ray in the $\hat x-\hat y$ plane, $\gamma$ is the angle the initial light ray velocity makes with the $\hat x$ axis. With these definitions, a choice of $\gamma$ and $p_z$ then fixes a particular light ray emanating from the origin. 

A wave front of null geodesics is a two parameter family of light rays. We construct this wave front by allowing the parameters $(p_z, \gamma)$ to run through all their allowed values, with the additional requirement of taking one set of values with the positive sign in both Eqns.~\ref{gamma} and one set of values with one positive and one negative sign.

\subsection{Time along null geodesics}

With the definitions for $x(s)$ in Eqn.~\ref{nullgeodesics} and $\dot t$ in Eqn.~\ref{tdot}, we can explicitly integrate to obtain a function for $t(s)$ with a boundary condition that $t = 0$ at $s=0$.  The result is rather long and complicated, but critically it is largely dominated by a linear term in $s$ with ``wiggles''.  The resulting time along null geodesics parametrized by $(p_z, \gamma)$ with $p_t = 1$ is

\begin{eqnarray} t(s) & = & \frac{1}{4 \alpha} \left\{ 2 \alpha (1+p_z^2) s + ( 1 - p_z^2) \left[ \sin{(2\gamma)} +  2 \sin{(\alpha_o s)} \right. \right. \nonumber \\ ~ & ~ & \left. \left. ~ + \sin{(2 (\gamma + \alpha_o s))} - 2 \sin{(2\gamma + \alpha_o s)} \right] \right\} \label{time}. \end{eqnarray}

\section{Individual Light Rays} \label{lightrays:sec}

We begin by considering individual light rays in the $\hat x - \hat y$ plane found by setting $p_z = 0$.  All such light rays form closed loops in the $\hat x - \hat y$ plane.

\begin{figure}
\begin{center}
\scalebox{0.6}{\includegraphics{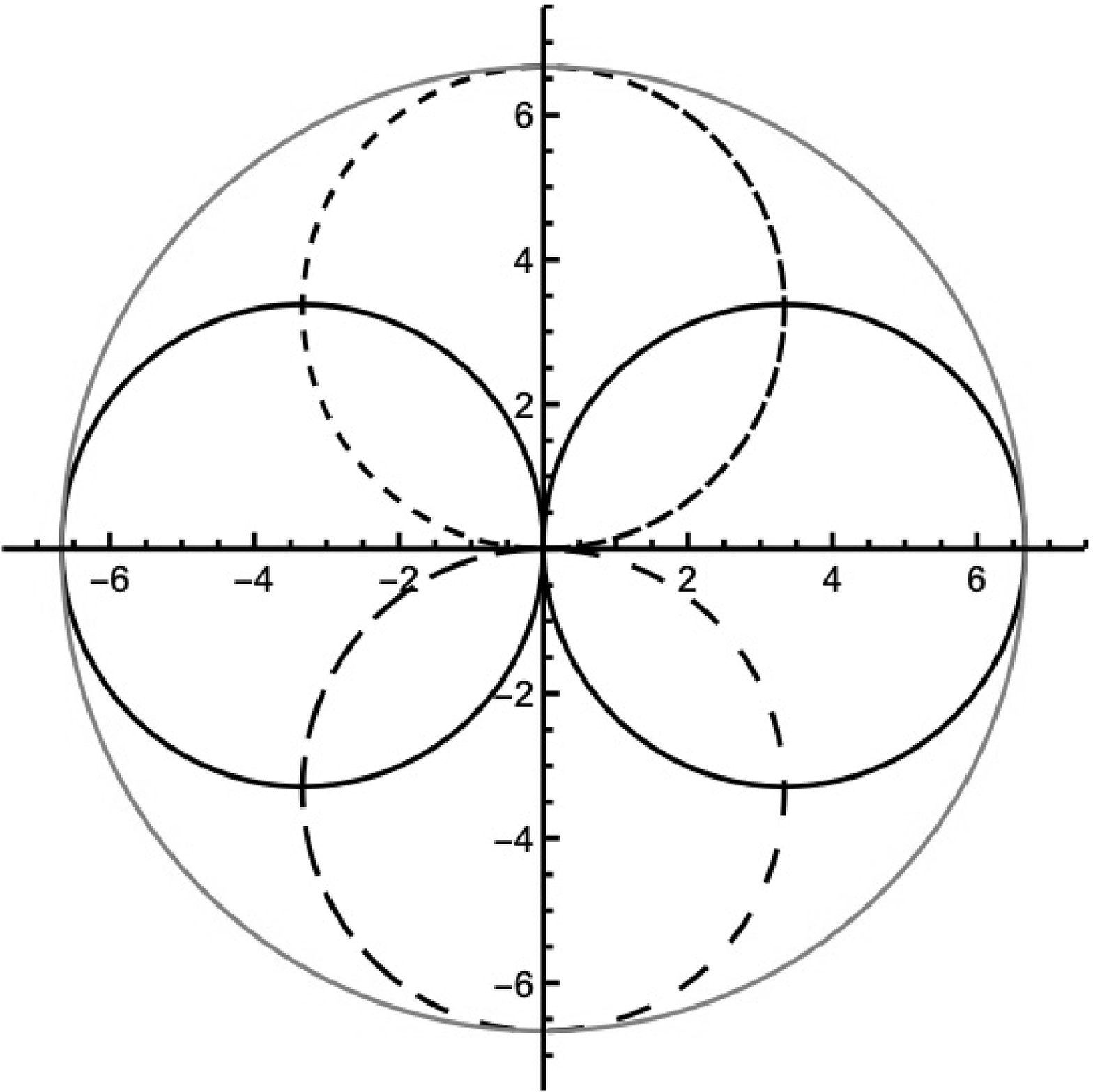}} \caption{\label{fourrays:fig}
Four lights rays in the $\hat x - \hat y$ plane for $p_z = 0$ and $\alpha_o = 0.3$ The large circle is a reference circle with a radius of $2/\alpha_o$.  The spatial part of each of the four light rays form closed loops. Rays for $p_y = -1$ (solid black on the right), $p_y=+1$ (solid black on the left), $p_y = 0$ with $v_{xo} < 0 $ (dashed, below $\hat x$ axis) and $p_y = 0$ with $v_{xo} > 0$ (dashed, above the $\hat x$ axis) are shown. } \end{center}\end{figure}

Figure~\ref{fourrays:fig} shows four light rays with varying values of $\gamma $ for Eqs.~\ref{nullgeodesics} and $\alpha_o = 0.3$.  We see that all rays form spatially closed loops that extend out to a maximum radius of $2/\alpha_o$.   We therefore will interpret the circle of radius $2/\alpha_o$ as the spatial part of a causal boundary beyond which an observer at the origin can not communicate directly.

These spatially closed light ray paths have complicated development in time.  We note that the $t$ coordinate represents the proper time of all observers who maintain a fixed position, and in particular, it is the proper time of a freely falling observer at the origin.  As we have seen, it is possible to form a system of freely falling observers at rest with respect to one another such that the proper time of all these observers is a set of synchronized clocks.

All light rays trace out circular paths that progress consistently in one direction (counterclockwise) in terms of the affine parameter $s$.  However, the behavior of light rays in terms of the time coordinate is more complicated.  For rays with $\gamma \approx 0$, the advancement of the time coordinate monotonically increases  with the affine parameter.  However, for larger $\gamma$, the time coordinate as a function of the affine parameter undergoes reversing periods where increases in affine parameter lead to decreases in the time.

This means that if an observer at the origin flashed a light in a particular direction, that it would trace a closed loop path through the space, but also that our system of observers at rest in the space-time would consider the light path to not circle around uniformly in time.  Rather, according to this system of observers, at a later time a pair of light pulses spontaneously appear (out of the blue-sky) at a point along this closed loop traveling in both directions along the loop as shown in Fig.~\ref{bluesky:fig}.  The part of the blue-sky pulse propagating with the original direction returns to the observer at the origin, while the portion of the pulse propagating in the other direction runs into the original pulse from the origin and the pair vanish.  Put another way, for some stationary observers, the light ray would appear to be traveling the the ``opposite'' direction from its propagation with respect to the affine parameter.

By comparison, the situation in the G\"odel metric for light rays in the azimuthal plane has some similarities and some striking differences.  Light rays that are emitted from the origin in the G\"odel space-time also form closed spatial loops in the azimuthal plane, with a distinct circular boundary that is taken as separating causal regions.  However, for all such light rays emitted from the origin in the G\"odel space-time, the time coordinate increases monotonically with the affine parameter.  

\begin{figure}
\begin{center}
\scalebox{0.8}{\includegraphics{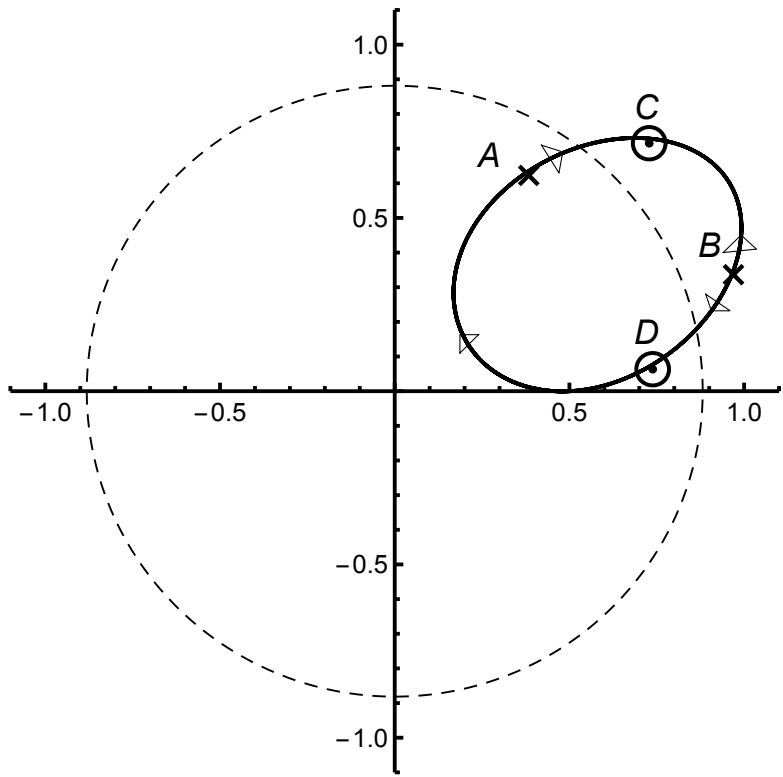}} \caption{\label{godelbluesky:fig}
A plot of the spatial path of a light ray in the $z=0$ plane in the G\"odel metric.  The light ray is emitted from $x=0.5$ and travels counterclockwise.  According to an observer at the origin, when the light ray has reached $A$ by moving directly from the initial point, a blue sky pair of pulses begins at $B$.  Part of this pulse travels back towards the original pulse and annihilates at $C$ with the initial pulse, when the forward moving pulse is at $D$.  Both the annihilation and blue sky positions are hidden from the observer at the origin due to the circle that delineates the causally connected region. } \end{center}\end{figure}

However, when considering light rays emitted from a point $x^a_p$, away from the origin in the G\"odel space-time, the time coordinate does not increase monotonically.  These rays, do however, cross over the circle that represents the causal boundary for an observer at the origin.  Similar to the case in the Type D metric, the mathematics indicates that the observer at the origin would consider the pulse emitted at $x^a_p$ to undergo the same type of behavior as described above, as is shown in Fig.~\ref{godelbluesky:fig}.  However, both the blue-sky appearance and the annihilation of the pulse pairs occurs outside the causal boundary for an observer at the origin, hiding much of the odd behavior \cite{roebuck}.

\section{Wave Fronts in the $\hat x - \hat y$ Plane} \label{planefronts:sec}

In this section, we begin to examine wave fronts of null geodesics emitted from a point source at the origin.  As each null geodesic advances in its affine parameter, it moves through the space as well as through time.  Our perspective is that an observer remaining stationary at the origin has emitted a pulse of light in all directions simultaneously.  We will consider points from different null geodesics to belong to the same wave front if they have the same $t$ coordinate - the proper time of the system of observers who remain at rest relative to the observer who released the light pulse at the origin.

\begin{figure}
\begin{center}
\scalebox{0.6}{\includegraphics{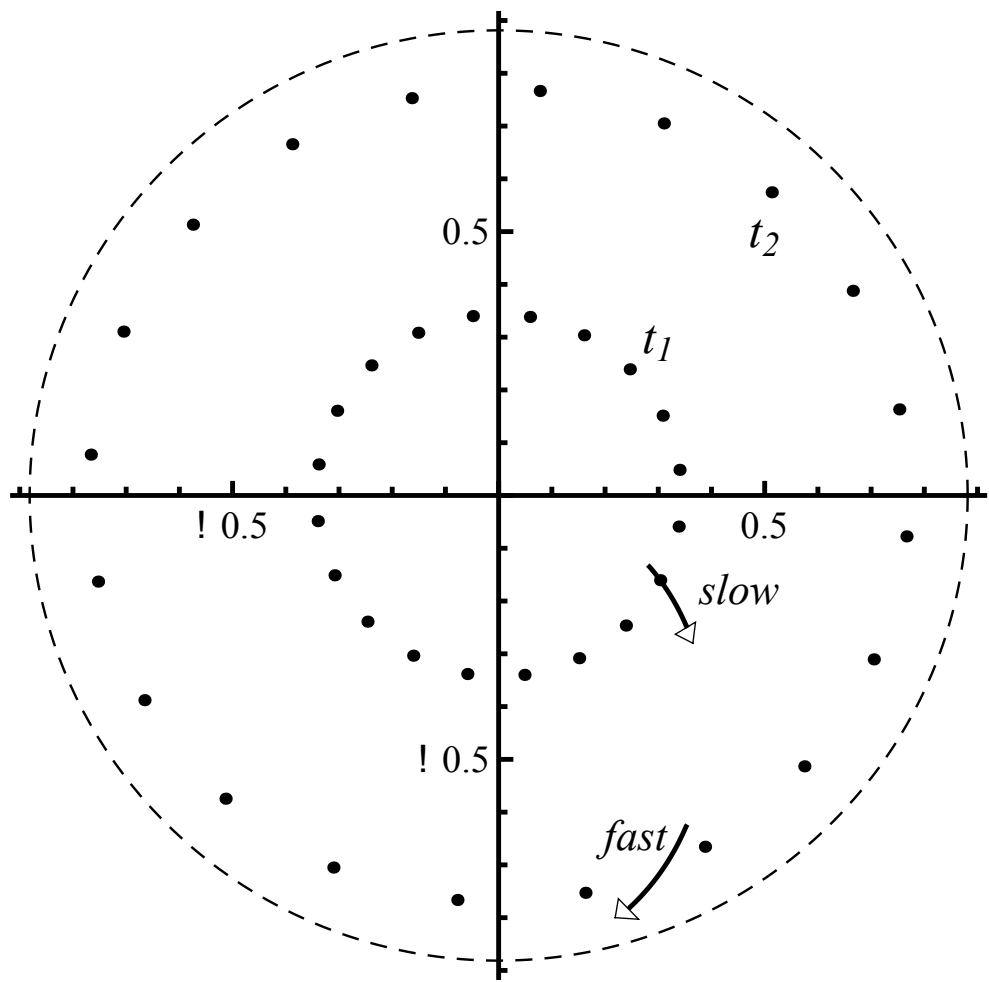}} \caption{\label{godelplane:fig}
Points on the wave front of null geodesics emitted from the origin in the G\"odel metric.  The wave fronts are circles that expand to the limiting circle and rebound.  The points on the wave front rotate demonstrating the G\"odel metric's rotational properties. } \end{center}\end{figure}

Because of conservation of momentum in the $\hat z$ direction, many of the key features of the wave fronts in G\"odel-type metrics can be understood from the advances of wave fronts in the $\hat x - \hat y$ plane.  In the case of a G\"odel metric, all the $\hat x - \hat y$ null geodesics emitted by an observer at the origin rebound off a circle that separates the causal region.  Figure \ref{godelplane:fig} shows points along the wave fronts in the G\"odel metric, which are rotating circles.

In the metric considered in this paper, the situation is significantly different, due to the time reversals revealed in Sec.~\ref{lightrays:sec}.  Figure~\ref{planefronts:fig} shows the wave front at eight consecutive times for a pulse of light emitted at the origin.  Due to the non-monotonic nature of the time coordinate along the null geodesics, wave fronts of constant time show ``blue sky'' style bifurcations in the $\hat x - \hat y$ plane where a portion of the wave front appears out of nowhere, attaches to the main wave front. Later a new section breaks off and disappears.  The entire wave front is rotating counterclockwise.  

Clear cusps appear on the wave front as well.  Cusps are a standard form of wave front singularity. We will see in the next section how these cusps arise as folds in the time cut functions of the light cone.

\begin{figure}
\begin{center}
\scalebox{0.45}{\includegraphics{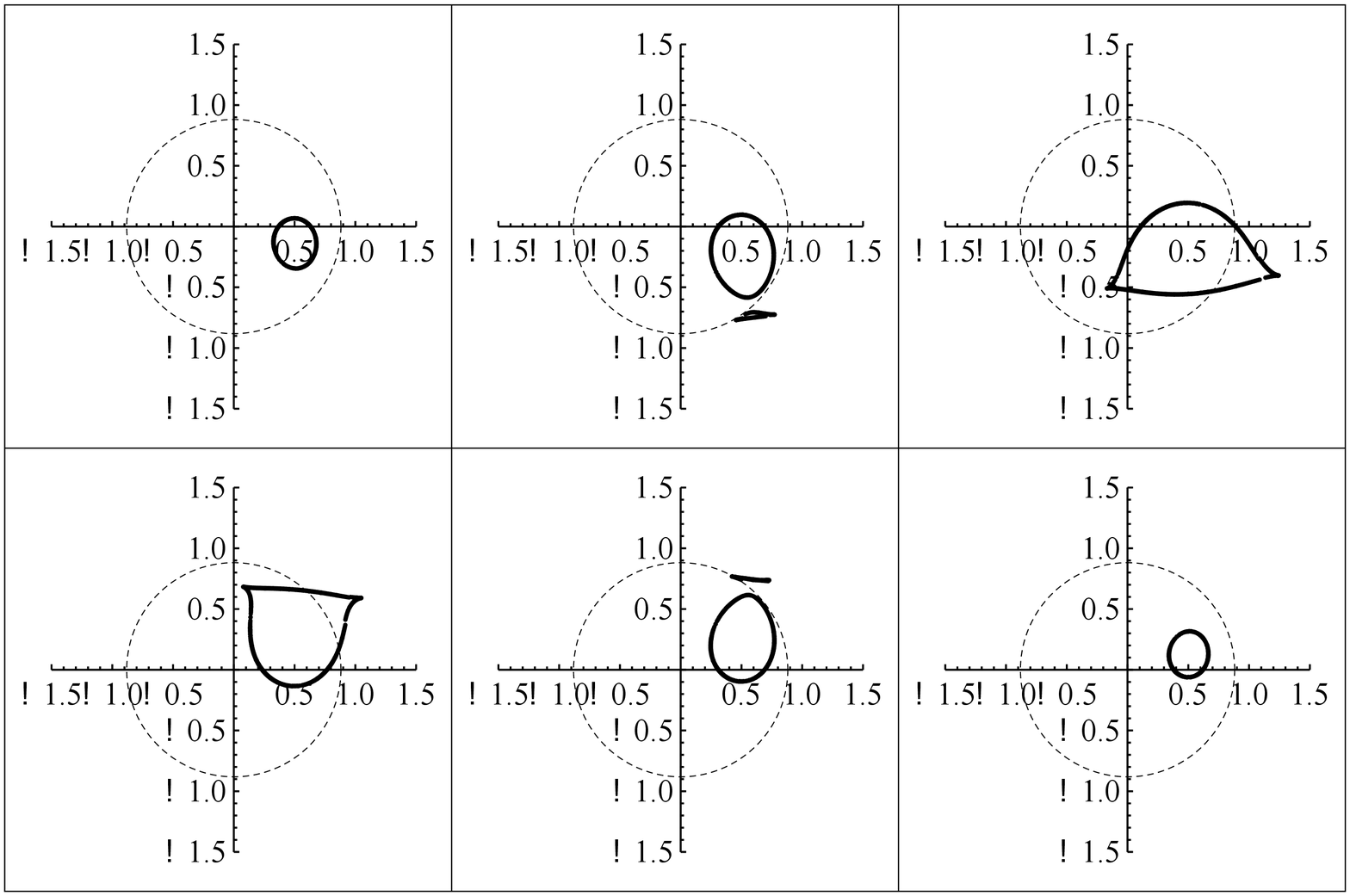}} \caption{\label{godelfront_notorigin:fig}
Points on the wave front of constant time for null geodesics emitted from a point away from the origin in the G\"odel metric.  The wave fronts develop blue sky bifurcations outside the circle for which the observer at the origin can communicate.  } \end{center}\end{figure}

The situation for wave fronts from the origin in the Type D metric is similar to that of the G\"odel metric for wave fronts of constant time constructed by an observer at the origin when the initial pulse of light is emitted away from the origin.  Figure~\ref{godelfront_notorigin:fig} shows the advancing wave front of constant time of a pulse of light in the $\hat x - \hat y$ plane  when the initial location of the pulse is away from the origin.  In this case, there are again blue sky bifurcations where portions of the wave front appear, then connect to the main portion, then break free and disappear.  

The principle difference between the wave fronts of constant time from points at the origin in the Type D metric and from points away from origin in the G\"odel metric is that the blue sky bifurcations in the G\"odel metric appear outside the circle that is the horizon for the observer at the origin.  This means that the observer at the origin who is re-constructing the wave-fronts using her time coordinate will never see or communicate with observers who saw disconnected portions of the wave front.  By contrast, the observer at the origin in the Type D metric would communicate with observers who indicate the presence of the portion of the wave front formed from the blue sky bifurcations.

\section{Time, Light Cones and Wave Fronts} \label{lightcones:sec}

\begin{figure}
\begin{center}
\scalebox{0.6}{\includegraphics{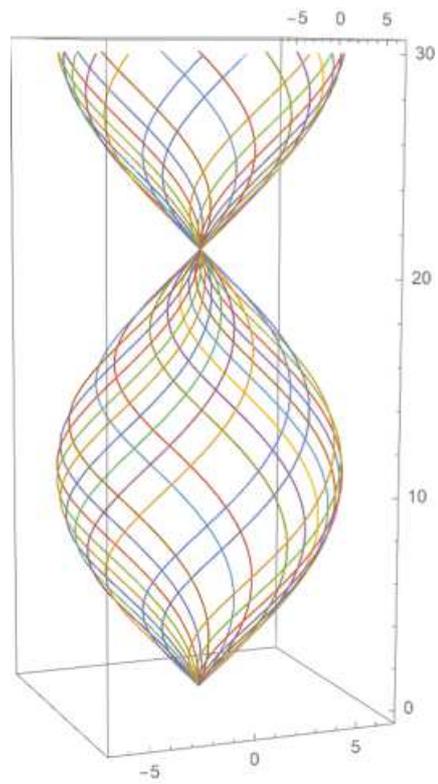}} \caption{\label{basiclightcone:fig}
The future light cone of a point at the origin suppressing the $\hat z$ direction. The $\hat x - \hat y$ positions of light rays with $p_z = 0$ are plotted against the affine parameter $s$ on the vertical axis. } \end{center}\end{figure}

As opposed to the constant time wave fronts, the light cones in the metric considered in this paper are very simple.  Figure~\ref{basiclightcone:fig} shows future light rays with $p_z = 0$ plotting the motion in the $\hat x- \hat y$ plane against the affine parameter $s$ on the vertical axis.  The light cone with the $\hat z$ components suppressed is simply a rotating series of light rays that rebound off the limiting circle of radius $2/\alpha_o$.

\begin{figure}
\begin{center}
\scalebox{0.6}{\includegraphics{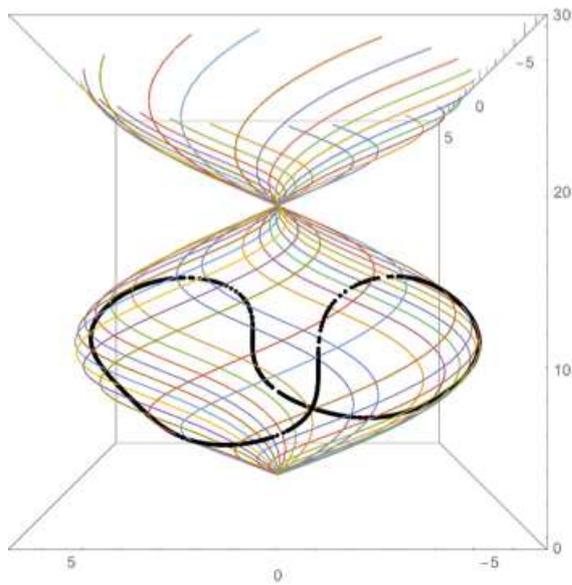}} \caption{\label{lightcone42:fig}
The future light cone of a point at the origin suppressing the $\hat z$ direction as in Fig.~\ref{basiclightcone:fig} showing points along light rays at $t=4.2$.  We see that the time cut folds over the light cone so that while the cut is one dimensional and smooth, its projection into the $\hat x - \hat y$ plane will have cusps. } \end{center}\end{figure}

The time coordinate along the light cone, however, is a complicated function.  Figure~\ref{lightcone42:fig} shows a series of reference light rays comprising the light cone and a large number of points along light rays when $t=4.2$ - which is a time where cusps have formed after a blue sky bifurcation in the time wave front - as we see in Fig.~\ref{planefronts:fig}.  This time cut folds over itself revealing cusp singularities as one sees in the canonical case presented in Arnol'd \cite{arnold1}.

\begin{figure}
\begin{center}
\scalebox{0.6}{\includegraphics{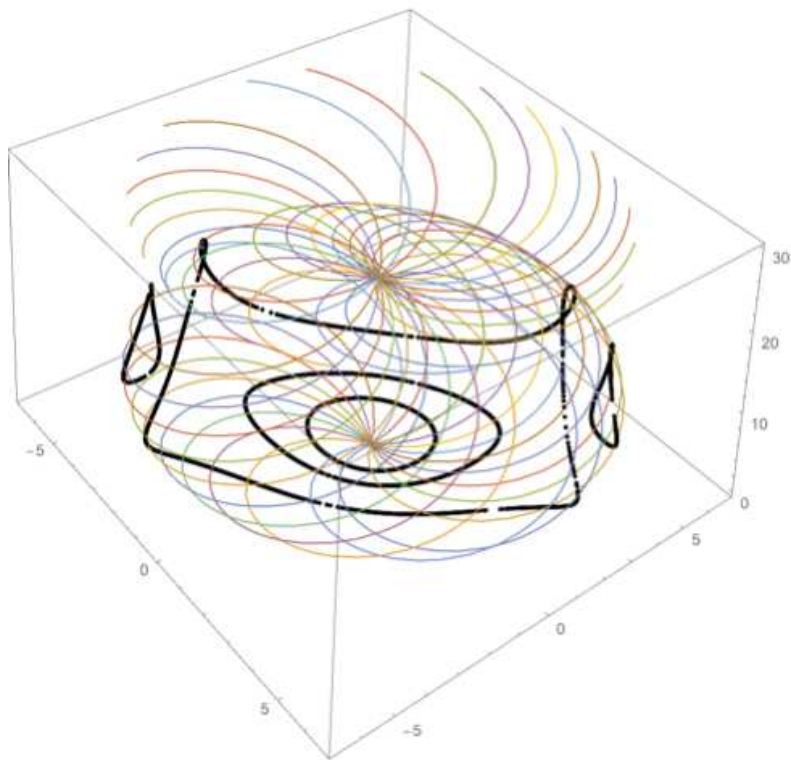}} \caption{\label{lightcone3:fig}
Three time cuts on the light cone at $t = 1.5$, $t = 2.4$ and $t = 4.2$.  The time cut at $t=1.5$ is the closest closed loop near the origin.  The cut at $t=2.4$ is in three parts due to the nature of the time coordinate along the light rays. } \end{center}\end{figure}

Figure~\ref{lightcone3:fig} shows three time cuts: at $t= 1.5$ as a small closed loop near the base of the light cone; at $t = 2.4$ which is a slightly larger loop, but also two separate side loops that have grown from blue sky bifurcations along the sides of the light cone; and finally at $t=4.2$ where the separate time cuts have rejoined the main group, but in a way that leads to overlaps that project into the $\hat x-\hat y $ plane with caustics.  This is made explicit in Fig.~\ref{lightcone3_above:fig} which shows the exact same light cone and three time cuts as Fig.~\ref{lightcone3:fig} but from directly above the light cone so that one sees the projection of the time cuts into the $\hat x- \hat y$ plane.

\section{Wave Fronts in Three Dimensions} \label{threeDfronts:sec}

The three-dimensional wave fronts of the Type D metric considered in this paper reveal how the caustic structures in the two-dimensional cases arise.  Figure~\ref{3dplots:fig} shows a series of six constant time wave fronts in three spatial dimensions for $\alpha = 0.3$.  Each plot is shown from the same vantage point with the $+x$ direction oriented out of the page and towards the right, the $+y$ direction is oriented towards the top of the page, and the $+z$ direction is oriented out of the page and towards the left.  The three-dimensional wave fronts begin as a topological sphere centered at the origin, but slowly distort along an axis in the $\hat x-\hat y$ plane.  

\begin{figure}
\begin{center}
\scalebox{0.55}{\includegraphics{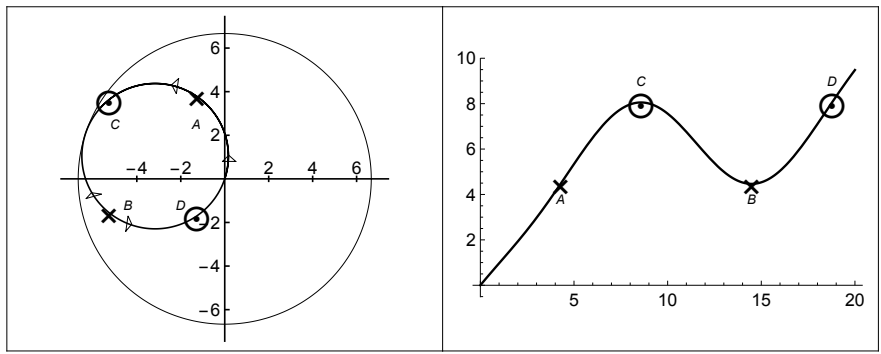}} \caption{\label{bluesky:fig}
On the left, a plot of the spatial path in the $\hat x - \hat y$ plane of a light ray with $\alpha = 0.3$, $p_y = 0.95$ and initial $p_x$ positive for $p_z=0$ where the light ray starts at the origin and moves counterclockwise.  On the right is a plot of the time coordinate for this null geodesic as a function of the affine parameter. According to an observer at the origin, at the point labeled $B$ a ``blue sky'' appearance of a pair of light pulses appears, moving along the closed curve in opposite directions.  At the point labeled $C$ one part of the pair meets up with the original pulse from the origin and the pair disappears. } \end{center}\end{figure}

\begin{figure}
\begin{center}
\scalebox{0.35}{\includegraphics{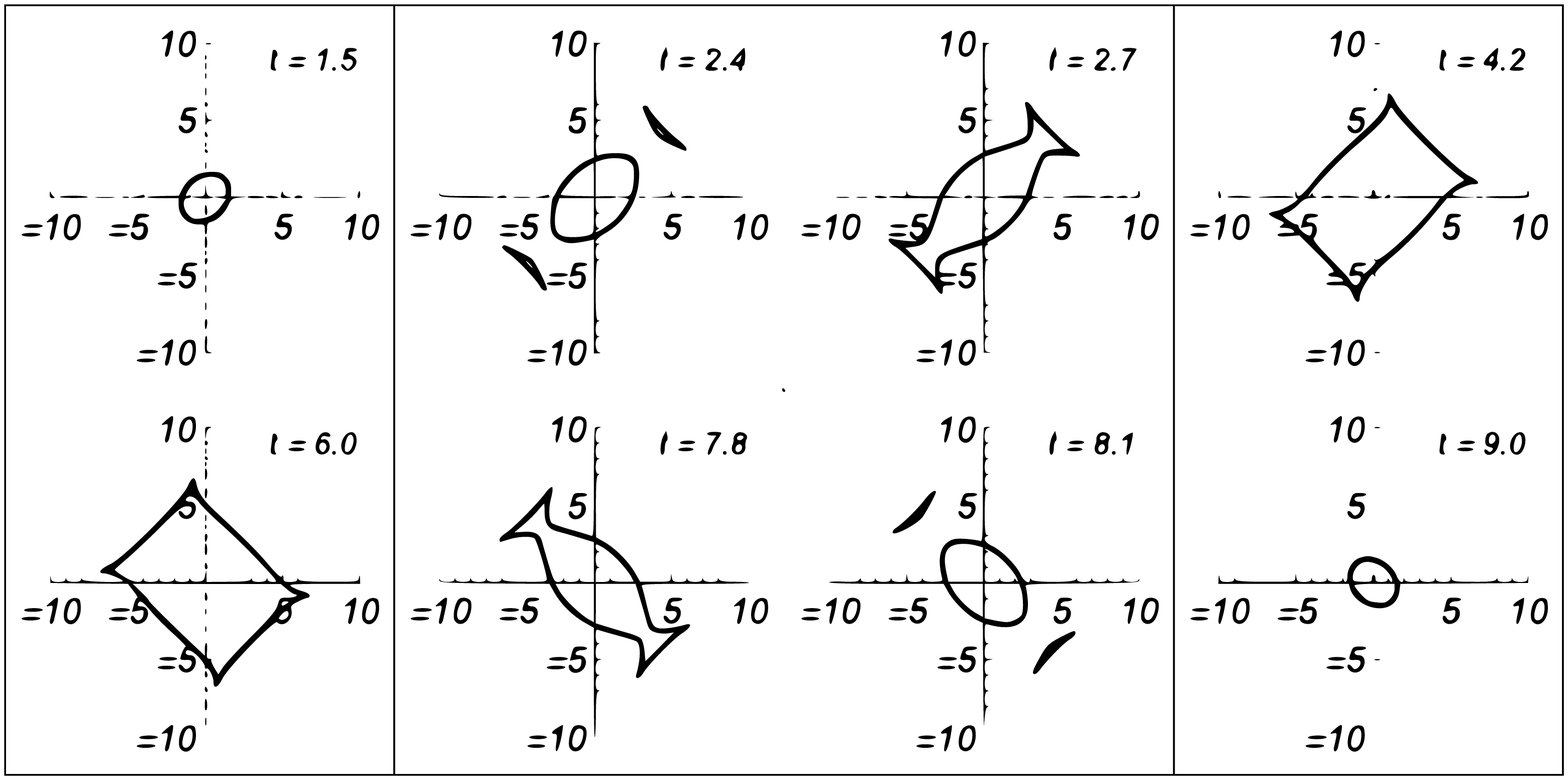}} \caption{\label{planefronts:fig}
Wave fronts of null geodesics emitted by an observer at the origin at eight successive times.  These wave fronts show bifurcations due to the non-causal nature of the time coordinate of the time coordinate shown in Fig.~\ref{bluesky:fig}.  Each entire wave front of constant time, including the disconnected sections, are inside the causal boundary circle of radius $2/\alpha_o$.} \end{center}\end{figure}

\begin{figure}
\begin{center}
\scalebox{0.6}{\includegraphics{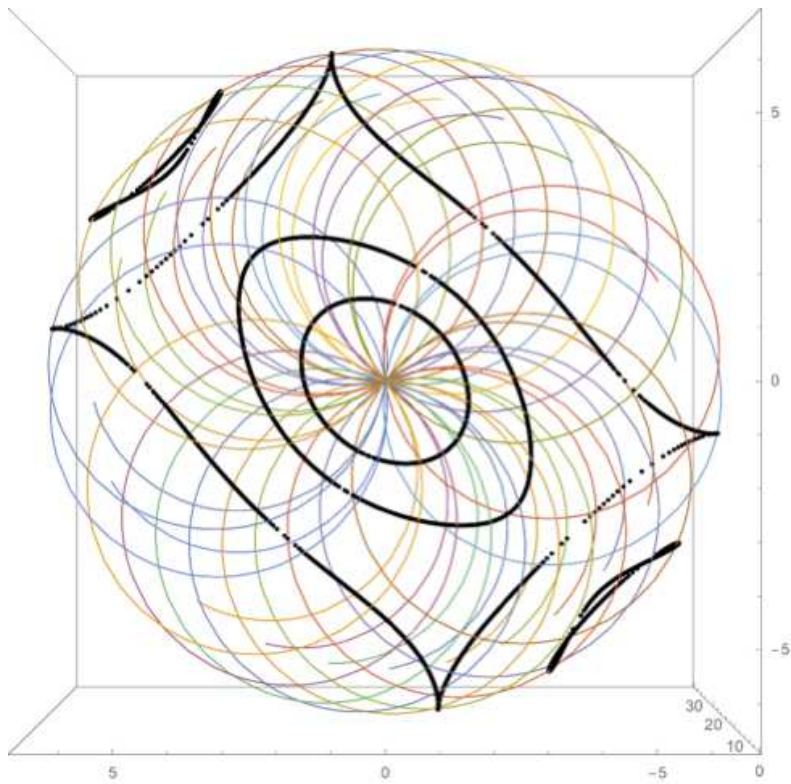}} \caption{\label{lightcone3_above:fig}
The same light cone and time cuts as in Fig.~\ref{lightcone3:fig}, but viewed from a point above the light cone along the affine parameter axis.  We see that the projection of these cuts is as shown in Fig.~\ref{planefronts:fig}. The projection of the outermost, disconnected portion is part of the wave front at $t=2.4$ causes a slight distortion in this figure away from the cusp shape seen in Fig.~\ref{planefronts:fig}. } \end{center}\end{figure}

\begin{figure}
\begin{center}
\scalebox{0.5}{\includegraphics{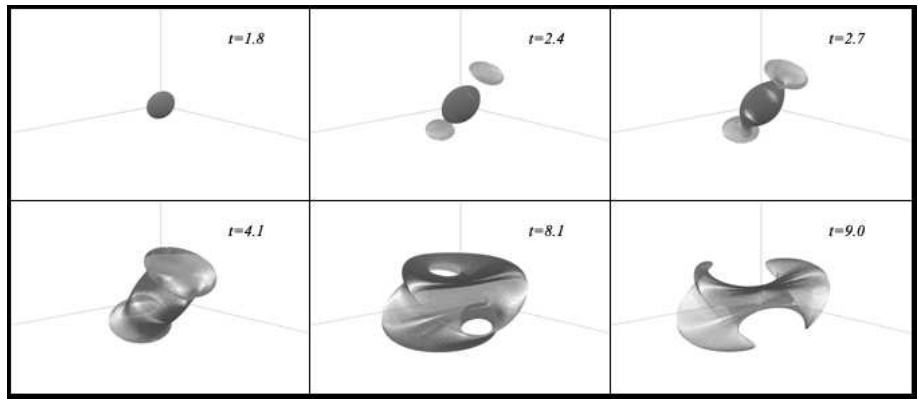}} \caption{\label{3dplots:fig}
Three dimensional wave fronts of null geodesics emitted by an observer at the origin at six successive times.  After an initial expansion, two distinct portions of the wave front appear separated from the main front at $t=2.4$.  These portions of the wave front rejoin the main branch shortly after, leaving two outer cusp ridges as the wave front singularities.  An opening forms before $t=8.1$ which leads to a pinching off of the wave front, which in Fig.~\ref{planefronts:fig} was the vanishing portion of the wave front in the plane.} \end{center}\end{figure}

\begin{figure}
\begin{center}
\scalebox{0.8}{\includegraphics{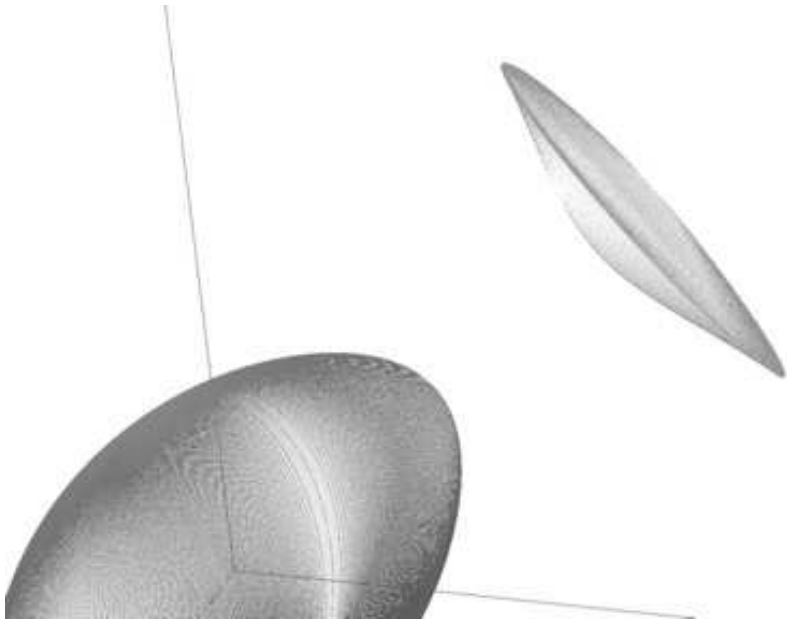}} \caption{\label{3dat2.4:fig}
A close view of a portion of the three-dimensional wave front at $t=2.4$.  The disconnected region displays a circular cusp ridge. } \end{center}\end{figure}

At $t=2.4$, we see a pair of blue sky bifurcations has developed in two areas.  We see in Fig.~\ref{3dat2.4:fig} that these blue sky sections have a three-dimensional structure with circular cusp ridge that bounds two smooth connecting surfaces.  The blue-sky bifurcation sections connect shortly after with the main wave front with two cusp ridges along the outside edge.  These parts of the wave front correspond to light rays being observed as in three spatial positions at a given time beginning at point $D$, as in Fig.~\ref{bluesky:fig}.

\begin{figure}
\begin{center}
\scalebox{0.55}{\includegraphics{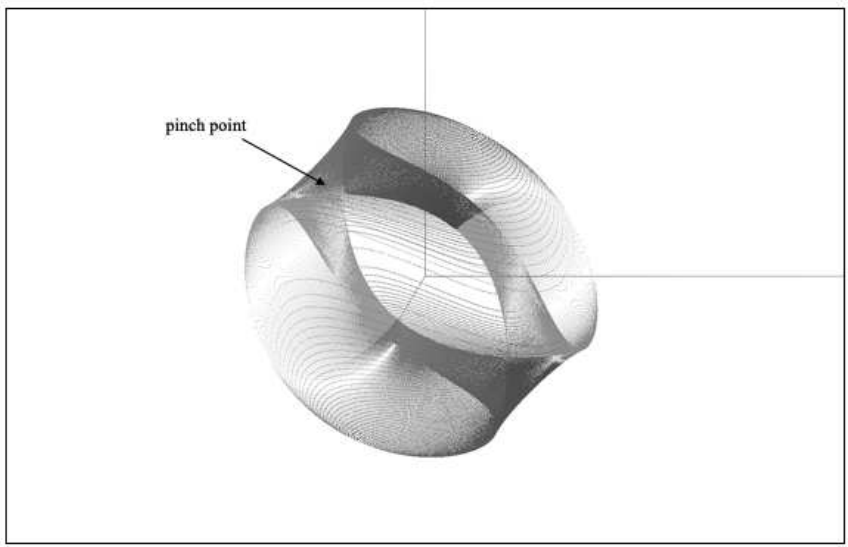}} \caption{\label{3d_pinch:fig}
The wave front at $t=7.95$ in the Type D metric.  We see that the wave front develops a pair of pinch points.  When these portions of the wave front come together, an openings appear and grow, eventually splitting the wave front into the elongated double spear-like structure seen in the last view in Fig.~\ref{3dplots:fig}. } \end{center}\end{figure}

Between roughly $t=2.7$ and $t = 7.8$, the wave front maintains a cusp ridge that rotates and begins to align with the $+z$ axis.  Near $t=7.95$ the wave front begins to collapse on itself as is shown in Fig.~\ref{3d_pinch:fig}.  In two regions symmetric with the $\hat z$ axis, the wave front comes together, and when the contact occurs, holes in the wave front become apparent as is shown in Fig.~\ref{3dplots:fig} at $t=8.1$.  Those holes quickly grow, and eventually the wave front breaks apart leaving a central tube, surrounding the $\hat z$ axis as it collapses and two spear or hook-like structures with cusp ridges that are advancing parallel to the $\hat z$ axis.  The breaking apart of the wave front in three-dimensions results in the blue-sky vanishing of a portion of the wave front in the $\hat x-\hat y$ plane in Fig.~\ref{planefronts:fig}.  These portions of the wave front are created when light rays which had three spatial positions at the same coordinate time according to our system of observers recombining at point $C$ in Fig.~\ref{bluesky:fig}.

\begin{figure}
\begin{center}
\scalebox{0.5}{\includegraphics{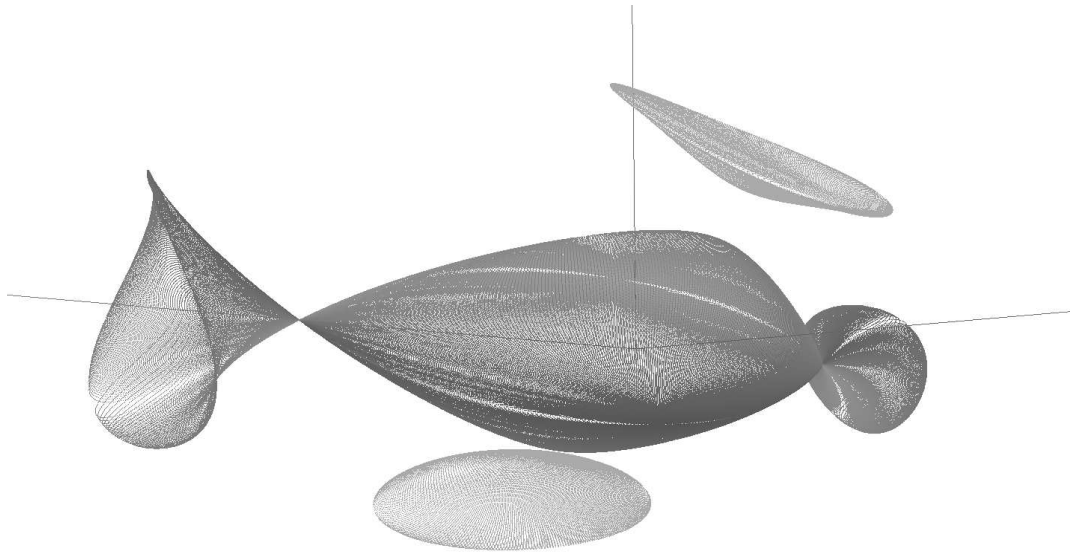}} \caption{\label{3d_at13:fig}
The wave front at $t=13.0$ in the Type D metric.  A new blue sky bifurcation has occurred with new sections of the null wave front appearing separated from the main wave front. We see cusp ridges on both ends of the wave front that are advancing from the spear in Fig.~\ref{3dplots:fig}. } \end{center}\end{figure}

\begin{figure}
\begin{center}
\scalebox{0.45}{\includegraphics{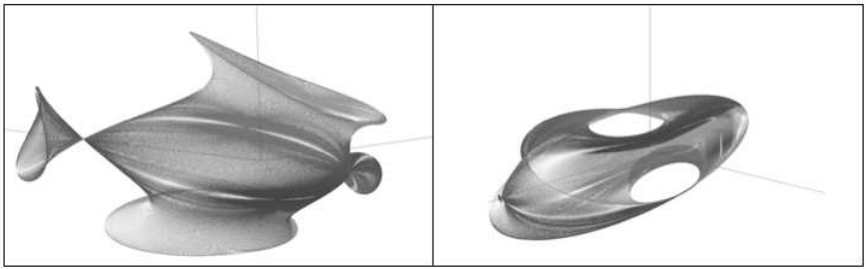}} \caption{\label{3d_late:fig}
The wave front at $t=14.0$ (left panel) and $t = 19.0$ (right panel). The separated portion of the wave front at $t=13$ has rejoined the main section.  This section rotates, collapses in pinch points, and eventually pulls apart as previously.} \end{center}\end{figure}

\begin{figure}
\begin{center}
\scalebox{0.4}{\includegraphics{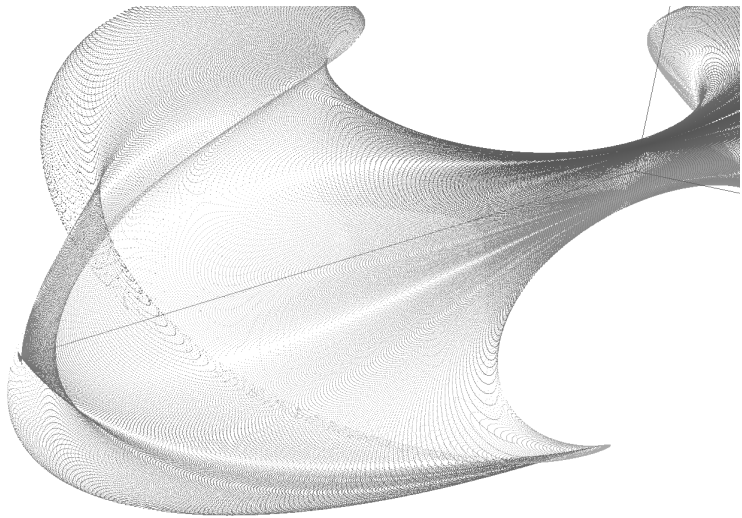}} \caption{\label{3d_at20:fig}
The wave front at $t=20.0$ shows a very small cusp ridge at the tip of a new spear structure.} \end{center}\end{figure}

The process continues as the light rays with new light rays undergoing time reversals.  In Fig.~\ref{3d_at13:fig}, we see that at $t=13$, a new disconnected region of the wave front appears.  These sections combine with the main body of the wave front as in Fig.~\ref{3d_late:fig}, and then the body rotates as before, develops two pinch points, and begins to open up. After the the opening breaks apart, a new set of spear-like sections appears, as is shown in Fig.~\ref{3d_at20:fig}.  A small cusp ridge remains on each end of the progressing wave front as a remnant of the previous spear structure.

\section{Discussion}

Wave fronts of light rays are fundamentally non-local reconstructions of a causal feature of any space-time.  The light cone itself is the central structure of causality; it separates regions of causal contact, and properties in the light cone structure control communication between points in the space-time. As we see in Figure~\ref{basiclightcone:fig}, the light cones in the space-time we study here are contained to a non-local region of the space-time but are not global structures extending throughout the entire space-time.

How one chooses to slice a light cone is not necessarily inherently obvious in a given space-time.  In some space-times, there may arise a natural system of observers who share some common notion of time.  For example, in the Friedman-Robertson-Walker (FRW) cosmological metrics, there is a natural, or preferred, frame of reference in which all the matter is homogeneous and isotropic, and this frame can be extended across the entire space-time.  Therefore, in FRW metrics, or metrics that are approximately FRW such as those considered in standard gravitational lensing studies, there is a natural slicing to the light cone associated with the clocks of observers at rest with respect to the cosmic flow.

On the other hand, in asymptotically flat space-times such as the Kerr-Newman metrics, there is not one system of observers following freely falling, or geodesic, trajectories that share a common synchronized clock system.  Even though these black-hole space-times are very physically motivated, general relativity does not permit a single global coordinate system of synchronized clocks with which to slice the light-cones.  The best one could do is to consider wave fronts of constant coordinate time, taking this time to physically mean the clock time of an observer in the asymptotically flat limit.

In Section~\ref{static_obs:sec}, we show that although the ``natural'' motion in the Type D metric considered in this paper is rotational, there is a family of observers who follow geodesic motions and remain at rest with respect to one another.  This family shares a common notion of time, and it is with respect to this system of observers that we have drawn our constant time wave fronts.  We note that a similar set of observers can be constructed in the G\"odel metric, though we do not show this result in this paper.

This result is surprising, as prior literature shows that there are no global surfaces of simultaneity in the Som-Raychaudhuri space-times. Section~\ref{static_obs:sec} shows that in the particular metric examined in this paper, that there is a set of geodetic observers who extend throughout the space-time region containing the light cone, who remain at rest with respect to one another, and who share a common notion of time.  In terms of these observers, who extend across a finite but non-local region, we plot wave fronts of constant time.

In the case of FRW metrics, null geodesic wave fronts emanating from a point source maintain spherical shapes.  While the wave fronts will remain spheres, the radius of those spheres as a function of the cosmic time coordinate does depend both on the expansion of the universe and the nature of space, for instance whether the universe is closed.  However, the wave front does not develop singularities, except possibly point-like singularities associated with the light traveling fully around a closed universe should that be possible before a big crunch.  Likewise, the wave front does not develop non-connected features, because while the light cones in FRW metrics might lead to particle horizons, there are no non-causal time features in the FRW metrics.

In this paper, we have shown that a central observer can organize and directly communicate with a system of observers who follow geodesic motion, share synchronized clocks, and are at rest with respect to one another.  When this central observer emits pulses of light in all directions, and reconstructs the wave fronts from her direct communication with a system of observers, non-connected wave fronts, and unusual wave front singularities are observed. 

The presence of non-connected regions in constant time wave fronts of light rays in the Type D metric considered in this paper is evidence of the strongly non-causal features of the space-time.  The fact that the wave fronts are not simply connected helps to demonstrate the notion that there is no globally defined surface of simultaneity. We strongly suspect that in a space-time where there is a globally defined surface of simultaneity, that it would not be possible to see non-connected regions in the constant time wave fronts of null geodesics.

Interestingly, in the G\"odel space-time, a similar set of observers (at rest with respect to the central observer, in direct communication, using a common sense of time) do not record non-causal features of the wave fronts from pulses of light emitted by the central observer.  For light emitted from the central position, no non-causal wave-front singularities are observed.

\section{Conclusions}

This paper examines wave front singularities in a G\"odel-type, Type D, metric.  The wave fronts demonstrate several interesting non-causal features of the metric, and the analysis of the metric and wave front bring to light interesting issues related to the study of wave fronts in gravitational lensing and the study of acausal metric spaces.

In terms of gravitational lensing, the spatial location of wave front singularities in the light cone of an observer are locations where significant magnification of sources of light appear.  In this paper, we see typical wave front singularities (such as the cusp ridge), as well as atypical features.  To our knowledge, the appearance of disconnected portions of the constant time wave fronts in regions where the observer receives light rays directly is unusual.  It does not, for instance, happen in the G\"odel metric.

The relation between the light cones and the wave fronts is also of interest.  The development of the light cones is smooth and consistent in terms of the affine parameter.  Figure \ref{lightcone3_above:fig} shows two examples of the formation of wave front singularities outlined in the literature, for instance in Arnol'd \cite{arnold1}.  First, in the connected portion at $t=4.2$, we see that time cut of the light cone is in one smooth portion along what Arnol'd calls the big wave front in space that includes the affine parameter, but that the projection of this cut into the space has cusps.  Second, we see an unusual feature of wave front singularities - that a disconnected portion of the wave front can be born away from the main section if the constant time cut of the light cone is not connected.

The appearance of this second form of null-geodesic wave front singularity distinguishes the G\"odel-type metric examined in this paper from the usual G\"odel metric.  These features indicate a broader class of acausal behavior than is typically seen, and the results of this paper further the understanding of the range of possible behavior permitted by general relativity, at least mathematically.

\section*{Acknowledgements}
Authors sincerely acknowledged the anonymous kind referees) for their valuable comments and suggestions which have greatly improved the present paper. ML acknowledges the financial support of Bridgewater State University through a Summer Grant from the Adrian Tinsley Program for Undergraduate Research.  The three-dimensional wave fronts plots were created using the Walnut package create by Eric Grotzke.  Light ray and wave front diagrams for the G\"{o}del space-time were originally reported in a paper by Thomas Kling, Kevin Roebuck, and Eric Grotzke \cite{roebuck}.

\section*{Data Availability}

No data associated with this manuscript.

\section*{Conflict of Interest}

Authors declare that there are no conflict of interest regarding publication this paper.

\pagebreak



\end{document}